\RequirePackage{fix-cm}
\documentclass[smallextended]{svjour3}       
\smartqed  
\usepackage{graphicx}
\begin{document}

\title{Ellipsoidal Expansion of the Universe, Cosmic Shear, Acceleration and Jerk Parameter}
%
%


\author{ Luigi Tedesco    }


\institute{Dipartimento di Fisica dell'Universit\`a di Bari, 70126 Bari, Italy
and 
INFN - Sezione di Bari, I-70126 Bari, Italy \\ \\
              \email{luigi.tedesco@ba.infn.it }           
}

\date{Received: date / Accepted: date}

\maketitle

\begin{abstract}
In order to study if the anisotropy of the spacetime may induce differences in the cosmic shear or in deceleration parameter we investigate in Bianchi I Universe the most general expression for the cosmic shear and we study the connections between deceleration parameter and cosmic shear.
We also consider a shear parametric approach  to measure the transition from decelerating to an accelerating Universe. We also study the connection between jerk parameter and ellipsoidal Universe.

\end{abstract}

\section{Introduction}
The standard model of the Universe is based on the assumption that it is homogeneous and isotropic. This is considered when we consider on a large scale. Of course when we consider "local scales" the Universe is not homogeneous and isotropic. But the Universe is "really" homogeneous and isotropic at large scales? This question is fundamental in cosmology, because when we assume  the homogeneity and isotropy we are assuming an "{\it priori}", because we forget that this is not always compatible with the large structure in the Universe.  This is outside the scope of this paper, but it is interesting to study this possibility, for example see \cite {{ehlers1},{Clarkson1},{Clarkson2},{russell}}, on the other hand we say that  we have important tensions between the standard model of the Universe and experimental data (see also \cite{{percival},{devega}}). In this paper we 
are considering a Bianchi I model of the Universe (that describe the Universe with the anisotropic and homogeneous background), in which we have two axis where the Universe expand itself. This is not an academic situation, because the last experimental data give us very "news" regards this not standard model of the Universe. There are several evidences as regards the anisotropy during the evolution of the Universe, for example the existence of anomalies in the CMB suggests that we have a plane-mirroring symmetry when we consider large scales \cite{{gurza1},{gurza2}}. Therefore the presence of small anisotropy deviations from the isotropy of the CMB is not of secondary importance, in fact we have small anisotropy with $10^{-5}$ amplitude. Another important observational evidence is connected with the large angle anomalies in CMB \cite{bennet}. 
As regards these anomalies 
we have four possibility to be considered: (1) the so called  "alignment of quadrupole and octupole
moments" \cite{{land},{raltston},{deoliveira},{copi}};
(2) the large-scale asymmetry \cite{{eriksen},{hansen}};
(3) a very strange cold spot \cite{vielva} and the low quadrupole moment in CMB \cite{{barrow2},{berera},{campanelli1},{campanelli2}} in which it is possible to indicate an ellipsoidal expansion of the Universe because the low quadrupole moment of CMB is suppressed at large scales.  This is very interesting because it possible to think at some unknown mechanisms or an anisotropic phase during the evolution of the Universe. It is important to stress that Bianchi I models can be considered alternatives to the standard FRW models but with (very) small deviations from the exact isotropy in order to explain the anisotropies and anomalies in the CMB and so on. This is in agreement with \cite{martinez}.  
\\
A very interesting connection between Bianchi I models and quadrupole, octupole and limit on the shear, vorticity, Weyl tensor has been studied in \cite{maart}.
\\
Recent observations SNIa \cite{{riess},{perlmutter}} and CMB anisotropy \cite{bennet} suggest that actual Universe is undergoing an acceleration expansion and $-1 \leq q \leq 0$, where $q$ is the so called "deceleration parameter". 
\\
In this paper we study the ellipsoidal Universe in relation to the cosmic shear and the deceleration parameter.  The deceleration parameter $q$ has a very important role to describe the expansion and the evolution of the Universe.  A  popular way to study is to consider q parametrized as a function of $a(t)$, the scale factor, or $z$ the redshift or $t$ the cosmic time \cite{{w1},{w2},{w3},{w4},{w5},{w6},{w7},{w8},{w9},{w10},{w11},{w12},{w13}}. The properties of this cosmic shear parametrization is obtained because we have considered the evolution of the Universe from a decelerated phase in which the matter dominates over dark energy to an accelerated expansion in which dark energy dominates over matter.
\\
The organization of the paper is as follows. In section 2 we describe a Bianchi I Universe, in section 3 we study the general expression for the cosmic shear $\Sigma$. In section 4 we study the ellipsoidal Universe in relation to the cosmic shear and the deceleration parameter in a decelerated and accelerated Universe. In section 5 we discuss the so called jerk parameter in relation to the cosmic shear and eccentricity of the ellpsoidal Universe. 
Finally in section 6 we have summarized the conclusions of this work. 
%
%
%
\section{Ellipsoidal Universe}
In this Section we are interested in Bianchi I metric with planar symmetry, in which the most general plane-symmetric line element is described by Taub line element 
\cite{taub}
\begin{equation}
\label{eq1}
d s^2 = dt^2 - a^2(t) (dx^2 + dy^2) - b^2(t) dz^2
\end{equation}
where $a$ and $b$ are the directional scale factors along the x,y and z axes respectively and they are functions of comic time only. If we have $a(t)=b(t)$  this reduces to the flat FLRW spacetime. The line element has spatial sections with planar symmetry (xy)-plane and an axis of symmetry along the z-axis, that is to say two equivalent longitudinal directions $x$ and $y$ and transverse direction $z$. The metric eq.(\ref{eq1}) is interesting because describes a space with an ellipsoidal expansion of the Universe at any moment of the cosmological time.
\\
The non-vanishing Christoffel symbols are
\begin{equation}
\Gamma^0_{11}=\Gamma^0_{22} = a \, \dot{a} \,\,\,\,\,\,\,\,\,\,\,\,\,\,\,\,\,\,\,\, \Gamma^0_{33}=b \, \dot{b}
\end{equation}
\begin{equation}
\,\,\, \Gamma^1_{01}=\Gamma^2_{02} = \dot{a}/a \,\,\,\,\,\,\,\,\,\,\,\,\,\,\,\,\,\,\,\, \Gamma^3_{03} = \dot{b}/b
\end{equation}
where we have used $\dot{ } \equiv d/dt$ the derivative with respect to the cosmic time. For completeness we also write the non-vanishing Ricci tensor components:
\begin{equation}
R^0_{0} = -2 \left( \frac {\ddot{a}} {a} + \frac {\ddot{b}} {b} \right)
\end{equation}
\begin{equation}
R^1_1 = R^2_2 = - \left[ \frac {\ddot{a}} {a} + \left( \frac {\dot{a}} {a} \right)^2 + \frac {\dot {a} \dot {b}} {a b} \right]
\end{equation}
\begin{equation}
R^3_3 = - \left( \frac {\ddot {b}} {b} + 2 \, \frac {\dot {a} \, \dot{b}} {a \, b} \right).
\end{equation}
The energy-momentum tensor consistent with planar symmetry is
\begin{equation}
T^{\mu}_{\,\, \nu} = {diag} \,  (\rho, -p_{\parallel}, -p_{\parallel} -p_{\perp})
\end{equation}
with $\rho$ energy density, $p_{\parallel}$ longitudinal pressure and $p_{\perp}$ transversal pressure. This energy-momentum tensor may be the sum of two components: anisotropic and isotropic contributions.
\\
Let us consider an anisotropic component in the Universe (for example domain wall, cosmic string, magnetic field), that induces the planar symmetry with an energy-momentum tensor 
\begin{equation}
(T_A)^{\mu}_{\,\, \nu} = {diag} \, (\rho^A, -p^A_{\parallel}, -p^A_{\parallel}, -p^A_{\perp})
\end{equation}
and an isotropic contribution (for example vacuum energy, cosmological constant, radiation, matter)
\begin{equation}
(T_I)^{\mu}_{\,\, \nu} = {diag} (\rho^I, -p^I, -p^I, -p^I)
\end{equation}
where it can be obtained by means of the three components: radiation (r), matter(m) and cosmological constant ($\Lambda$):
\begin{equation}
\rho^I= \rho_r + \rho_m+ \rho_{\Lambda}
\end{equation}
\begin{equation}
p^I= p_r + p_m + p_{\Lambda}
\end{equation}
with $p_r=\rho_r/3, \,\,\, p_m=0, \,\,\, p_{\Lambda} = \rho_{\Lambda}$. In \cite{{berera1},{berera2},{barrow}} the authors studied the exact solutions of the Einstein's equations with different plane-symmetric and isotropic components.
\\
We define 
\begin{equation}
A\equiv (a^2 b)^{1/3}, \,\,\,\,\,\,\,\,\, H_a= \frac {\dot{a}} {a}, \,\,\,\,\,\,\,\,\, H_b= \frac {\dot{b}} {b}
\end{equation}
with $A$ the average scale factor of the plane symmetric metric, $H_a$ and $H_b$ the directional Hubble's parameters in the direction of  x,y and z respectively.  
The average expansion rate or "mean Hubble parameter",  takes the form
\begin{equation}
H= \frac {\dot{A}} {A}.
\end{equation}
The proper volume $V$ is defined as
\begin{equation}
V = \sqrt{ -g} = a^2b \, .
\end{equation}
It is possible to write $H$ in terms of $H_a$ and $H_b$:
\begin{equation}
H=\frac {\dot {A}} {A} = \frac {\frac {d} {dt} [(a^2b)^{1/3}]} {(a^2 b)^{1/3}} = \frac {2 H_a +H_b} {3}.
\end{equation}
The expansion scalar $\theta$ is
\begin{equation}
\theta = 3H = 3 \frac {\dot {{A}}} {A}  ,
\end{equation}
the scalar shear is
\begin{equation}
{\sigma}^2 = \frac {1} {2} \sigma_{ij} \sigma^{ij} =\frac {1} {2}  \left( 2 H_a^2 + H_b - \frac {\theta^2} {3} \right)
\end{equation}
and the deceleration parameter $q$ is 
\begin{equation}
\label{deceleration}
q= - \frac {A \ddot{A}} {{\dot{A}}^2} = \frac {d} {dt} \left( \frac {1} {H} \right) -1.
\end{equation}
The cosmic shear vector is $\vec{\Sigma}$, whose components are
\begin{equation}
\Sigma_{x,y,z} \equiv \frac {H_{x,y,z} -H} {H}\, .
\end{equation}
It is also possible to write
\begin{equation}
\Sigma_a \equiv \frac {H_a - H} {H} = \frac {H_a -H_b} {2 H_a +H_b}
\end{equation}
with 
\begin{equation}
2 \, \Sigma_a + \Sigma_b=0.
\end{equation}
For semplicity we set $\Sigma_a \equiv \Sigma $.
\\
Let us define the eccentricity of the Universe
\begin{equation}
e= \sqrt{1 - \frac {b^2} {a^2}}  \phantom{sddhwei} a>b   \phantom{sddiiwei} {or}  \phantom{sddhwwei} e= \sqrt{1 - \frac {b^2} {a^2}}  \phantom{sdhdwei} b>a
\end{equation}
in which the first case applies for example for magnetic field and the second for domain wall. The eccentricity is normalized in order to have today, at the present cosmic time,  $a(t_0)=b(t_0)=1$.
In order to obtain the most general evolution equation for the eccentricity of the Universe, we start by Einstein's equations, in Bianchi I Universe:
\begin{eqnarray}
\left\{
\begin{array}{rl}
&
\left(\frac {\dot{a}} {a} \right)^2 + 2 \frac {\dot{a} \,  \dot{b}}  {a \, b} = 8 \pi G \,(\rho^I + \rho^A) \\ \\
&
\frac {\ddot{a}} {a} + \frac {\ddot{b}} {b} + \frac {\dot{a} \,  \dot{b}}  {a \, b} = - 8 \pi G \, (p^I + p^A_{\parallel}) \\ \\
&
2 \frac {\ddot{a}} {a} + \left(\frac {\dot{a}} {a} \right)^2 = -8 \pi G \, (p^I + p^A_{\perp})
\end{array}
\right.
\end{eqnarray}
that may be write in terms of $H_a$ and $H_b$:
\begin{eqnarray}
\label{eqdiH}
\left\{
\begin{array}{rl}
&
H_a^2+ 2H_a H_b = 8 \pi G \rho \\ \\
& 
{\dot{H}_a} + H_a^2 + {\dot{H}}_b + H^2_b + H_a H_b = - 8 \, \pi \, G \, p_{\parallel} \\ \\
&
2 {\dot{H}}_a + 3 H_a = - 8 \pi = - 8 \, \pi \,  G \, p_{\perp}.
\end{array}
\right.
\end{eqnarray}
where $\rho=\rho^I + \rho^A$, \, $p_{\parallel} = p^I +p_{\parallel}^A$ and \, $p_{\perp} = p^I + p_{\perp}^A$.
\\
It is possible to connect $H_a$ and $H_b$ to the eccentricity, in fact for the case $e= \sqrt{1- b^2/a^2}$ we have $a^2(1-e^2)=b^2$. Deriving the two sides with respect to the time we have $ 2\, b \, \dot{b}  = 2 \, a \, \dot{a} (1-e^2) -2 a^2 e \, \dot{e} $ and dividing for $b^2$ we obtain
\begin{equation}
\label{eqHb} 
H_b=H_a - \frac  {e \, \dot{e}} {1 - e^2}
\end{equation}
while in the second case, for $e=\sqrt{1-a^2/b^2}$ we have
\begin{equation}
\label{eqHa}
H_a=H_b - \frac  {e \, \dot{e}} {1 - e^2}.
\end{equation}
In this way it is possible to write the shear $\Sigma$ in terms of the eccentricity, remembering that $\Sigma=H_a/H-1$:
\begin{equation}
\label{sigmaa1}
\Sigma = \frac {e \, \dot{e}} {3 H (1-e^2)}  \phantom{ sddhdiiwiwei} {for} \phantom{ sddhdiiwiwei} e= \sqrt{1-b^2/a^2}
\end{equation}
and
\begin{equation}
\label{sigmaa2}
\Sigma = - \frac {e \, \dot{e}} {3 H (1-e^2)}  \phantom{ sddhdiiwiwei}  {for}  \phantom{ sddhdiiwiwei} e= \sqrt{1-a^2/b^2}.
\end{equation}
In order to obtain the most general temporal evolution for the eccentricity, let us subtract the second equation from the third in eq.~(\ref{eqdiH}):
\begin{equation}
\label{ecc}
\frac {d} {dt} \left( \frac {e \, \dot{e}} {1 - e^2}  \right) + 3 H \left( \frac {e \, \dot{e}} {1 - e^2} \right) = \pm \,  8 \pi G \, (p_{\parallel} - p_{\perp})
\end{equation}
where "+" refers to the eccentricity $e = \sqrt{1-b^2/a^2}$ and the sign "-" refers to $e = \sqrt{1-a^2/b^2}$.
\\
It is important to obtain the evolution equation for the anisotropic energy density $\rho^A$. In order to this end, we remember that the anisotropic component of the energy-momentum  tensor is conserved 
\begin{equation}
\label{cons}
(T^A)^\mu_{\,\, \nu; \mu} =0
\end{equation}
that, considering the $\nu=0$ component, gives:
\begin{equation}
\label{consrho}
{\dot{\rho}}^A + 2 \, H_a (\rho^A + p^A_{\parallel}) + H_b \, (\rho^A + p^A_{\perp}) =0.
\end{equation}
It is very interesting and useful to see how scales $\rho^A$. We observe that 
\begin{equation}
\label{HHa} 
H_a= H \pm \frac {e \, \dot{e}} {3 (1-e^2)}
\end{equation}
\begin{equation}
\label{HHb} 
H_b= H \mp \frac {2} {3} \frac  {e \, \dot{e}} { (1-e^2)}
\end{equation}
where the sign "+" and "-" in eq.(\ref{HHa}) refers to $e=\sqrt{1-b^2/a^2}$ and $e=\sqrt{1-a^2/b^2}$ respectively (similarly for the eq.(\ref{HHb})). Inserting eqs.(\ref{HHa}) and (\ref{HHb}) in eq.(\ref{consrho}) we have
\begin{equation}
\label{rhopunto}
{\dot{\rho}}^A + 3 H \rho^A + 2 H p^A_{\parallel} + H p^A_{\perp} \pm \frac {2} {3} p^A_{\parallel} \frac {e \, \dot{e}} {1-e^2} \mp \frac {2} {3} p^A_{\perp} \frac {e \, \dot{e}} {1-e^2} =0.
\end{equation}
Now we assume a barotropic relation between pressure and density, therefore the equations of state are 
$p^A_{\parallel} = w^A_{\parallel} \rho_A$ and $p^A_{\perp}=w^A_{\perp} \rho^A$, that inserted in eq.(\ref{rhopunto}) and neglecting $p^A_{\parallel} - p^A_{\perp} \simeq 0$, we have the following differential equation:
\begin{equation}
{\dot{\rho}}^A + H \rho^A \, (3+2 w^A_{\parallel} + w_{\perp})=0, 
\end{equation}
therefore
\begin{equation}
\label{rhosol}
\rho^A= \rho^A_{(0)} \frac {1} {A^{3+2w^A_{\parallel}+w^A_{\perp}}}
\end{equation}
where $\rho^A_{(0)}$ is the anisotropic energy density now. Eq.(\ref{rhosol}) represents the direct connection between anisotropic energy density and the equation of state; the connection is obtained by means the terms $w^A_{\parallel}$ and $w_{\perp}$ (see also \cite{{pippo1},{pippo2}}.
\section{General expression for the cosmic shear $\Sigma$}
It is possible to connect the cosmic shear $\Sigma$ with the eccentricity, in order to obtain the most possible information and to discuss in appropriate way the physical   consequence in relation to these parameters that are a good indication about the ellipsoidal expansion of the Universe. The interesting thing is that these lines of reasoning may be obtained in any temporal frame of the evolution of the Universe, therefore if we know the actual value of $\Sigma$ and $e$ (=1) we are able to obtain the value of the shear connected with the value of the eccentricity in any moment of the developmental time position of the Universe. What we want to find is a general equation that connects $\Sigma$ and $e$ for any time. Let us consider  the evolution equation eq.(\ref{ecc}), and we set $y\equiv H(t) \, \Sigma(t)$. In this way we obtain the following differential equation
\begin{equation}
\frac {dy} {d t} + 3 Hy= \frac {8 \pi G} {3} (p^A_{\parallel} - p^A_{\perp}).
\end{equation} 
The general solution of this equation is
\begin{equation}
y(t) = \frac{1} {A^3} \left[\frac {8 \pi G} {3} \int dt' (w^A_{\parallel} - w^A_{\perp}) \rho^A A^3 + c \right] 
\end{equation}
with $c$ integration constant.  
\\
Putting 
$
f(A) \equiv \frac {8 \pi G} {3} \int dt' (w^A_{\parallel} - w^A_{\perp}) \rho^A A^3 
$
we have
\begin{equation}
y(t) = \frac {1} {A^3} [f(A) + c] .
\end{equation}
We remember that $A_0 \equiv A(t=0)=1$, therefore the integration constant is
$c=y(t_0)-f(A_0)$
and the general  solution is
\begin{equation}
y(t) = \frac {1} {A^3} \, [f(A)+y(t_0) -f(A_0)].
\end{equation}
Putting all together, we have 
\begin{equation}
\label{putting}
\Sigma =  \frac {H_0} {A^3 H} \left[\Sigma^{(0)} + g(A)\right]
\end{equation}
where 
\begin{equation}
\bar{w} = \frac {1} {3} (2 w_{\parallel}^A + w^A_{\perp})
\end{equation}
is the mean equation of state parameter, $g(A)$ given by 
\begin{equation}
g(A) = \Omega^{0}_A (w_{\parallel} - w_{\perp} )\int_1^A \frac {du} {u^{ 1+3 \bar{w}} H(u)/H_0}
\end{equation}
and
\begin{equation}
\Sigma^0 = \Omega^0_A (w^A_{\parallel} - w^A_{\perp} \int_0^1 \frac {du} { u^{1+3 \bar{w}} \, H(u)/H_0}
\end{equation}
We assume that our components are not interacting. Let us introduce density parameters 
\begin{equation}
\Omega_{X} = \frac {\rho_X} {\rho_c} \;\;\;\;\;\;\;\;\; \rho_c = \frac {3 H^2} {8 \pi G}
\end{equation}
with $X=r, m, \Lambda$ (dark energy), $A$
and the relation
\begin{equation}
\Omega_r+\Omega_m+\Omega_{\Lambda} + \Omega_A=1.
\end{equation}
where of course
\begin{equation}
\frac {H(A)} {H_0} = \sqrt{ \Omega_R A^{-4} + \Omega_m A^{-3} + \Omega_{\Lambda} + \Omega_A A^{-3 -2 w^A_{\parallel}-w^A_{\perp} }}
\end{equation}
If the Universe isotropize (the discussion is developed in \cite{{campanelli1},{campanelli2}}, we must consider the limit for $A \rightarrow 0$. Taking into account eq. (\ref{putting}) we have 
\begin{equation}
\label{putting2}
\Sigma = \frac {\Omega^0_A (w^A_{\parallel} - w^A_{\perp})} {A^3 \, H/H_0 } \int_0^A \frac {du} {u^{1+3 \bar{w}} \, H/H_0}.
\end{equation}
Let us consider a real situation, for example, radiation era, where 
\begin{equation}
\frac {H} {H_0} = \sqrt{\Omega_R \, A^{-4}}
\end{equation}
In this way we have for the shear the expression
\begin{equation}
\Sigma= \frac {\Omega^0_A (w^A_{\parallel} - w^A_{\perp})} {\Omega_R (-2 w^A_{\parallel} - w^A_{\perp})} \, A^{1-2 w^A_{\parallel} - w^A_{\perp}}
\end{equation}
As regards this last equation,  simple analysis tell us that  
\begin{equation}
{lim}_{A \rightarrow 0} \,\,\, \Sigma =0  \phantom{ sddhdiiwiwei}  {if}  \phantom{ sddhdiiwiwei} \bar{w}  < 1/3  
\end{equation}
and 
\begin{equation}
\label{siggma}
{lim}_{A \rightarrow 0} \Sigma =- \frac {\Omega^0_A} {\Omega_R}  (w^A_{\parallel} - w^A_{\perp})  \phantom{ sddhdiiwiwei} {if}  \phantom{ sddhdiiwiwei} \bar{w}= 1/3
\end{equation}
that is not zero and this is an interesting result that must be considered when we take into account the parameter of the cosmological model of the Universe. It is important to observe that the shear may be positive or negative in eq.~(\ref{siggma}) and this will be important when we discuss in the next section the deceleration parameter of the Universe with the cosmic shear.
%
%
%
\section{Deceleration parameter and cosmic shear}
A very important parameter in cosmology is the so called "deceleration parameter" $q$ that is an interesting quantity that describes the nature of the expansion and evolution of the Universe and it is defined as
\begin{equation}
\label{deceleration}
q= - \frac {{A \ddot {A}}} {{{\dot{A}}^2}} = \frac {d} {dt} \left(\frac {1} {H} \right) -1.
\end{equation}
where $A$ is the average scale factor.
When we consider cosmological models we need to understand about the evolving Universe from decelerating phase ($q<0$) to actual accelerating phase ($q>0$).
Of course the experimental informations tell us that we are in a accelerated phase of the Universe, on the other hand many indicators go in this direction. It is possible to translate all these in terms of the $q$ parameter, in fact we can say the in the accelerating phase it is necessary that $-1<q <0$. Therefore it is very interesting to connect $q$ to the shear and eccentricity $e$. Let us consider eq.(\ref{sigmaa1}) for simplicity (the other case is the same), in other terms it is possible to write the expression for $H= e \dot{e}/ [3 \Sigma( 1- e^2)]$. The temporal derivative for $H$ is
\begin{equation}
\dot{H} = - \frac {\dot{\Sigma} \, e \, \dot{e}} {3 \, \Sigma^2 (1-e^2)} + \frac {1} {3 \Sigma} \left[ \frac {{\dot{e}}^2 + e^2 {\dot{e}}^2 + e \ddot{e} - e^3 \ddot{e}} {(1-e^2)^2}\right] \, ,
\end{equation}
in this way it is possibile, by eq.~(\ref{deceleration}), to obtain the most general expression that connects $q$, $\Sigma$ and $e$:  
\begin{equation}
\label{deceleration2}
q=-1 - \frac {3 (1-e^2) \Sigma} {e^2} - 6 \, \Sigma + \frac {3 (1-e^2) \dot{\Sigma}} {e \, \dot{e}} - \frac {3 (1-e^2) \Sigma \, \ddot{e}} {2  \, e^2 \, \dot{e}}
\end{equation} 
It is interesting to consider the connection between the evolution of the Universe starting from an ellipsoidal expansion and accelerated Universe. In literature often we study at what "z" redshift starts the acceleration of the Universe, in connection with different parameters, for example the study of the so called red shift drift and so on. Now we want to connected this physical situation with the eccentricity, the cosmic shear and the acceleration of the Universe. Let us consider eq.~(\ref{deceleration2}) that we can simplify observing that the eccentricity is small (for example at CMB time \cite{{campanelli1},{campanelli2}} $e \simeq 10^{-2}$), $1- e^2 \simeq 1$,  moreover $\ddot{e} \simeq 0$ because we can suppose that the variation of velocity of $e$ in time is very slow. Without lose of generality we suppose that $\dot{\Sigma}=0$, that is to say the variation in time of the cosmic shear is slow. Putting all together, eq.~(\ref{deceleration2}) becomes 
\begin{equation}
\label{dec3}
q= -1 - \frac {3 \, \, \Sigma} {e^2}
\end{equation}
A large number of recent cosmological observations strongly  suggest that the present Universe is accelerating in the present epoch \cite{{riess},{perlmutter}}. This introduce the so called dark energy in which we parametrize all  we don't know. But when the dark energy is subdominant we must consider a decelerated phase in order to have for example the formation of structure on large scale. In other terms when we consider the evolution of the Universe we must consider both cases in which decelerated and accelerated Universe must be considered in different epochs. 
The condition of decelerated Universe is $q>0$, therefore eq.~(\ref{dec3}) gives  the following condition:
\begin{equation}
\label{qminore0}
\Sigma <0 \,\,\,.
\end{equation}
Today the acceleration of the Universe requires that the parameter "q"
lies between $-1 <q_0<0$, that is to say, the condition on the cosmic shear is:
\begin{equation}
\label{dec4}
-\frac {e^2} {3} < \Sigma <0 \,\,. 
\end{equation}
It is interesting to note that for both cases (decelerated and accelerated Universe) the shear is negative, but if we consider an ellipsoidal Universe the permitted values are  in a very small interval between $-e^2/3$ and $0$. Moreover eq.~(\ref{qminore0}) indicates that in a decelerated expansion of the Universe the shear condition is independent from the eccentricity. On the other hand if we consider a Bianchi I Universe in the early stage, the cosmic shear does not assume all negative values but only permitted by eq.~(\ref{dec4}). Instead a decelerate Universe is able to permit  all negative values of the cosmic shear, with or without ellipsoidal expansion of the Universe.  
%
%
%
\section {Jerk parameter and ellipsoidal Universe}
In cosmology we don't know the value of $a(t)$ in relation to the entire history of the Universe. Therefore we consider the actual value of $a(t)$ and its derivative (Hubble parameter, deceleration parameter and so on). Taking tis into account we can extract limited information about the relevant physics of the Universe. To this end let us consider the Taylor expansion of the scalar factor around our time $t_0$ (the subscript "0" indicates that the coefficients are considered at $t=t_0$. 
\begin{equation}
a(t) = a_0  \left[    1  + H_0 (t-t_0) - \frac {1} {2} q_0 H_0^2 (t - t_0)^2 + \right. \nonumber \\
+ 
\frac {1} {3!} j_0 H_0^3 (t-t_0)^3 \left. + \frac {1} {4!} s_0 H_0^4 (t-t_0)^4 +...\right] \, .
\end{equation}
The deceleration parameter is 
\begin{equation}
q(t) = - \frac {a \, \ddot{a}} {{\dot{a}}^2} = \frac {d} {dt} \left(\frac {1} {H} \right) -1 \, ,
\end{equation} 
the jeark parameter is
\begin{equation}
\label{jeark}
j(t) = \frac { {d^3 a/ dt^3}} {a \, H^3} = q + 2q^2 - \frac {\dot q} {H}
\end{equation}
and the snap parameter is 
\begin{equation}
s(t) = \frac { {d^4 a/dt^4}} {a \, H^4} \, .
\end{equation}
$H(t)$ has the dimension of inverse of time, while $q(t)$, $j(t)$ and $s(t)$ are dimensionless parameters. The expression that involves the fifth and sixt derivatives of the parameter $a(t)$ are called {\it crackle} and {\it pop} respectively. The origin of the  names "snap, crackle, pop" goes back to the advertisement in 1932  of  "kellog's Rice Krispies" with "merril snap, crackle and pop in a bottle of milk". The jerk has been studied but with no very intensity 
\cite{Visser:2003vq,Poplawski:2006na,Dunajski:2008tg,Zhai:2013fxa,Luongo:2013rba,Dunajski:2014xha,Dantas:2015zhy,Mukherjee:2016trt}
parameter of our $\Lambda {CDM}$ flat model of the Universe is 1, therefore is a convenient quantity to consider in relation to models close to  $\Lambda {CDM}$. In other terms it is not academic information to consider jerk parameter for alternative models of Universe. Our Universe has the transition deceleration-acceleration, this is obtained for models with $j_0>0$ and $q_0<0$. Moreover it is important to say that when we consider jerk and snap coefficient, they are related to the third and fourth terms in the Taylor series expansion of the Hubble law, therefore it is very difficult to measure them because they are connected with the third and four derivative od the scale factor respectively. in this context the extraction of the information from these data of course depends on the accuracy of the experimental measures. The core of the problem is that in cosmology we don't know the value of $a(t)$ in relation to the entire history of the Universe.
\\
Let us consider eq.(\ref{jeark}) and considering the following positions as in the case of the acceleration parameter, $1- e^2 \simeq 1$ and $\ddot{e}=0$, we obtain the expression for the jerk parameter:
\begin{equation}
\label{jcorr}
j= 1+ 18 \, \Sigma + 72\,  \Sigma^2 + 9 \, \frac {\Sigma} {e^2} + 54 \, \frac {\Sigma^2} {e^2} + 9 \, \frac {e \,  \Sigma^2} {{\dot {e}}^3} \, .
\end{equation} 
Let us suppose to consider only the values that go as $1/e^2$ in eq.(\ref{jcorr}) (eccentricity is small) therefore we have
\begin{equation}
j= 1 + \frac {9 \, \Sigma} {e^2} \, (1 + 6 \Sigma).
\end{equation}
A very interesting reconstruction of the dark energy equation of state
by means a parametrization of the jerk parameter has been
very recently given by Luongo  \cite{Luongo:2013rba}, in which
we rewrite the jerk parameter as 
\begin{equation}
j = 1 + \epsilon
\end{equation}
with $\epsilon >0$ in this way it is possible to identify 
\begin{equation}
\epsilon = \frac {9 \, \Sigma} {e^2} \, (1 + 6 \Sigma).
\end{equation}
The presence of $\epsilon$ is a representation of the departure from the standard cosmology. 
This simple position is very interesting, because we can discuss the possibility to obtain $\Lambda$CDM model, that is to say $j=1$, when $\epsilon=0$. We have two cases.
The first is 
\begin{equation}
\Sigma = - \frac {1} {6}
\end{equation}
that is to say when the shear has this value we have the standard $\Lambda$CDM universe. 
\\
The second case is more intriguing because we re-obtain  $\Lambda$CDM universe when $\Sigma=0$. In both cases it is not considered the value of the eccentricity $e$. But when we consider $\Lambda$CDM universe we also say that the eccetricity is zero. Therefore in the second case it is necessary to consider the limit
\begin{equation}
\label{rere}
\lim_{e \rightarrow 0} \, \frac {\Sigma} {e^2} =0
\end{equation}
In other words $\Sigma$ must be an infinitesimal of superior order with respect to $e^2$, that is to say $\Sigma$ must go to zero as fast as possibile in relation to $e^2$. This is an intriguing result because generally speaking we isotropize putting $\Sigma \rightarrow 0$ without considering that in Bianchi I model of the Universe this procedure  is connected with the resetting of the eccentricity. We have in the denominator $e^2$ therefore is delicate to put $ e=0$ without considering $\Sigma =0$. In other terms is the coexistence of $\Sigma$ and $e^2$ in eq.(\ref{rere}) that give us the possibility to re-consider $\Lambda$CDM model starting from an ellipsoidal Universe. 
%
\section{Conclusions}
In the present work we have considered an ellipsoidal Universe and we have studied the most general expression of  cosmic shear in this Bianchi I Universe. We have connected cosmic shear with deceleration parameter in this model of Universe. We have considered the physical case in which the Universe transits from early decelerating phase to actual phase accelerating phase. In this paper we  have analyzed the transition phase of $q$ when deceleration to acceleration occurs at transition age. The two cases has been studied in which has been considered the cosmic shear as indicator of our discussion. We have found that cosmic shear must be negative for decelerated and accelerated Universe, but the permitted values for an accelerated Universe are very restricted. This may be an intriguing indicator as regards the possibility to have an actual Bianchi I Universe, that is to say it possibile to exclude it but  if we have a very very small actual eccentricity, the negative cosmic shear may be the signal of a Bianchi I actual Universe.  On the other hand we have considered that the variation of the acceleration, the jerk parameter, may be connected with eccentricity and cosmic shear in order to have more informations about these parameters. 
%
%
%


\begin{thebibliography}{}
\bibitem{ehlers1} J. Ehlers, P. Geren, and R. K. Sachs, J. Math. Phys. (N.Y.) {\bf 9},
1344 (1968).
%
\bibitem{Clarkson1} C. A. Clarkson and R. Barrett, Classical Quantum Gravity
{\bf 16}, 3781 (1999).
%
\bibitem{Clarkson2} C. Clarkson and R. Maartens, Classical Quantum Gravity
{\bf 27}, 124008 (2010).
%
\bibitem{russell} 
  E.~Russell, C.~B.~Kilinc and O.~K.~Pashaev,
  Mon.\ Not.\ Roy.\ Astron.\ Soc.\  {\bf 442}, no. 3, 2331 (2014).
%
\bibitem{percival} W. Percival, Mon. Not. R. Astron. Soc. {\bf 381}, 1053 (2010).
%
\bibitem{devega} H. J. de Vega, M. C. Falvella, and N. G. Sanchez,
arXiv:1009.3494.
%
%
\bibitem{gurza1} V. G. Gurzadyan, A. A. Starobinsky, T. Ghahramanyan, A.
L. Kashin, H. G. Khachatryan, H. Kuloghlian, D. Vetrugno,
and G. Yegorian, Astron. Astrophys. {\bf 490}, 929 (2008).
%
\bibitem{gurza2} V. G. Gurzadyan, T. Ghahramanyan, A. L. Kashin,
H. G. Khachatryan, A. A. Kocharyan, H. Kuloghlian, D.
Vetrugno, and G. Yegorian, Astron. Astrophys. {\bf 498}, L1
(2009).
%
\bibitem{bennet} C. L. Bennet, R. S. Hill, G. Hinshaw, M. R. Nolta, N.
Odegard, L. Page, D. N. Spergel, and J. L. Weiland,
Astrophys. J. Suppl. Ser. {\bf 148}, 97 (2003).
%
\bibitem{land} K. Land and J. Magueijo, Phys. Rev. Lett. {\bf 95}, 071301 (2005).
%
\bibitem{raltston} J. P. Ralston and P. Jain, Int. J. Mod. Phys. D {\bf 13}, 1857 (2004).
%
\bibitem{deoliveira} A. de Oliveira-Costa, M. Tegmark, M. Zaldarriaga, and A.
Hamilton, Phys. Rev. D {\bf 69}, 063516 (2004).
%
\bibitem{copi} C. J. Copi, D. Huterer, and G. D. Starkman, Phys. Rev. D {\bf 70},
043515 (2004).
%
\bibitem{eriksen} H. K. Eriksen, F. K. Hansen, A. J. Banday, K. M. Gorski, and
P. B. Lilje, Astrophys. J. {\bf 605}, 14 (2004); 609, 1198 (2004).
%
\bibitem{hansen} F. K. Hansen, A. J. Banday, and K. M. Gorski, Mon. Not. R.
Astron. Soc. {\bf 354}, 641 (2004).
%
\bibitem{vielva} P. Vielva, E. Martinez-Gonzalez, R. B. Barreiro, J. L. Sanz,
and L. Cayon, Astrophys. J. {\bf 609}, 22 (2004).
%
\bibitem{barrow2}  J. D. Barrow, Phys. Rev. D {\bf 55}, 7451 (1997).
%
\bibitem{berera} A. Berera, R. V. Buniy, and T.W. Kephart, J. Cosmol.
Astropart. Phys. {\bf 10}, 016 (2004).
%
\bibitem{pippo1} L. Campanelli, P. Cea, G.L. Fogli and L. Tedesco, Int. J. of Mod. Phys. D {\bf 20}, 1153 (2011).
%
\bibitem{pippo2} L. Campanelli, P. Cea and L. Tedesco, Mod. Phys. Lett. A {\bf 26}, 1169 (2011). 
%
\bibitem{campanelli1} L. Campanelli, P. Cea, and L. Tedesco, Phys. Rev. Lett. {\bf 97},
131302 (2006).
%
\bibitem{campanelli2} L. Campanelli, P. Cea, and L. Tedesco, Phys. Rev. D {\bf 76},
063007 (2007).
%
\bibitem{martinez} E. Martinez-Gonzales and J.L. Sanz, A. and A. {\bf 300}, 346 (1995).
%
\bibitem{maart} R. Maartens, G.F.R. Ellis, W.R. Stoeger, A. and A., {\bf 309}, L7 (1996).
%
\bibitem{riess} A.G. Riess {\it et al.} Astron. J. {\bf 116}, 1009 (1998).
%
\bibitem{perlmutter} S. Perlmutter {\it et al}. Astrophys. J. {\bf 517}, 565 (1999).
%
%
\bibitem{bennet} C.L. Bennet {\it et al.} Astrophys. J. Suppl. {\bf 148}, 1 (2003). 
%
\bibitem{w1} M.S. Turner and A.G. Riess, Astrophys. J. {\bf 569}, 18 (202).
%
\bibitem{w2} A.G. Riess {\it et al.} Strophys. J. {\bf 607}, 665 (2004).
%
\bibitem{w3} Y. Gong and A. Wang, Phys. Rev. D {\bf 73}, 083506 (2005).
%
\bibitem{w4} Y. Gong and A. Wang, Phys. Rev. D {\bf 75}, 043520 (2007).
%
\bibitem{w5} J.V. Cunha and J.A.S. Lima, MNRAS {\bf 390}, 210 (2008).
%
\bibitem{w6} J.V. Cunha, Phys. Rev. D {\bf 79}, 0473301 (2009).
%
\bibitem{w7} B. Santos {\it et al}. arXiv:1009:2733 [astro-ph.CO].
%
\bibitem{w8} R. Nair {\it et al} JCAP {\bf 01} 018 (2012).
%
\bibitem{w9} O. Akarsu {\it et al}. EPJ Plus {\bf 129}, 22 (2014).
%
\bibitem{w10} L. Xu and H. Liu, Mod. Phys. Lett A {\bf 23}, 1939 (2008).
%
\bibitem{w11} L. Xu and H. Liu, Mod. Phys. Lett A {\bf 24}, 369 (2009).
%
\bibitem{w12} S. del Campo {\it et al}. Phys.Rev D {\bf 86}, 083509 (2012).
%
\bibitem{w13} A.A. Mannon and S. Das, Int. J. Mod. Phys. D {\bf 25}, 1650032 (2016).
%
\bibitem{taub} A.H. Taub, Ann. Math. {\bf 53}, 472 (1951).
%
\bibitem{berera1} A. Berera, R.V. Buniy and T.W. Kephart, J. Cosmol. Astropart. Phys. {\bf 10}, 016 (2004).
%
\bibitem{berera2} R.V. Buniy, A. Berera and T. Kephart, Phys. Rev. D {\bf 73}, 063529 (2006).
%
\bibitem{barrow} J.D. Barrow, Phys. Rev. D {\bf 55}, 7451 (1997).
%
\bibitem{Visser:2003vq} 
  M.~Visser,
  Class.\ Quant.\ Grav.\  {\bf 21}, 2603 (2004).
%
\bibitem{Poplawski:2006na} 
  N.~J.~Poplawski,
  Phys.\ Lett.\ B {\bf 640}, 135 (2006).
\bibitem{Dunajski:2008tg} 
  M.~Dunajski and G.~Gibbons,
  Class.\ Quant.\ Grav.\  {\bf 25}, 235012 (2008).
\bibitem{Zhai:2013fxa} 
  Z.~X.~Zhai, M.~J.~Zhang, Z.~S.~Zhang, X.~M.~Liu and T.~J.~Zhang,
  Phys.\ Lett.\ B {\bf 727}, 8 (2013).
\bibitem{Luongo:2013rba} 
  O.~Luongo,
  Mod.\ Phys.\ Lett.\ A {\bf 28}, 1350080 (2013).
\bibitem{Dunajski:2014xha} 
  M.~Dunajski,
  Gen.\ Rel.\ Grav.\  {\bf 46},  1814 (2014).
%
\bibitem{Dantas:2015zhy} 
  C.~C.~Dantas and A.~L.~B.~Ribeiro,
  Phys.\ Rev.\ D {\bf 93},  043509 (2016).
\bibitem{Mukherjee:2016trt} 
  A.~Mukherjee and N.~Banerjee,
  Phys.\ Rev.\ D {\bf 93},  043002 (2016).


\end{thebibliography}


\end{document}